\documentclass[referee]{raa}     
\pdfoutput=1      
\usepackage{graphicx,times}
\usepackage{natbib}
\usepackage{amssymb,amsmath}
\bibpunct{(}{)}{;}{a}{}{,}
\graphicspath{{}}

\usepackage[a4paper=true,driverfallback=dvips,pagebackref=true]{hyperref}
\hypersetup{pdftitle = The title of my PDF, pdfauthor = My name, pdfsubject= The subject, pdfkeywords = keyword1 keyword2 keyword3} 
\hypersetup{colorlinks = true, linkcolor = green, anchorcolor = red, citecolor = blue, filecolor = red, pagecolor = red, urlcolor = red}
\usepackage[pagebackref=true]{hyperref}

\renewcommand{\vec}[1]{ {\mathbf #1} }

\newcommand{\Fig}{{Figure}}

\begin{document}

   \title{A New Approach of Data-driven Simulation and Its Application to Solar Active Region 12673   $^*$
\footnotetext{\small $*$ Supported by the National Natural Science Foundation of China.}
}

 \volnopage{ {\bf 20XX} Vol.\ {\bf X} No. {\bf XX}, 000--000}
  \setcounter{page}{1}

   \author{Zhi-Peng Liu\inst{1}, Chao-Wei Jiang\inst{1}, Xin-Kai Bian\inst{1}, Qing-Jun Liu\inst{1}, Peng Zou\inst{1}, Xue-Shang Feng\inst{1}
   }

   \institute{ Institute of Space Science and Applied Technology, Harbin Institute of Technology, Shenzhen, China; {\it chaowei@hit.edu.cn} \\   
\vs \no
   {\small Received 20XX Month Day; accepted 20XX Month Day}
}

\abstract{The solar coronal magnetic field is a pivotal element in the study of eruptive phenomena, and understanding its dynamic evolution has long been a focal point in solar physics. Numerical models, driven directly by observation data, serve as indispensable tools in investigating the dynamics of the coronal magnetic field. This paper presents a new approach to electric field inversion, which involves modifying the electric field derived from the DAVE4VM velocity field using ideal Ohm's law. The time series of the modified electric field is used as a boundary condition to drive a MHD model, which is applied to simulate the magnetic field evolution of active region 12673. The simulation results demonstrate that our method enhances the magnetic energy injection through the bottom boundary, as compared with energy injection calculated directly from the DAVE4VM code, and reproduce of the evolution of the photospheric magnetic flux. The coronal magnetic field structure is also in morphological similarity to the coronal loops. This new approach will be applied to the high-accuracy simulation of eruption phenomena and provide more details on the dynamical evolution of the coronal magnetic field.
\keywords{Sun: cornal mass ejections (CMEs) --- Sun: magnetic fields --- Methods: numerical 
}
}

   \authorrunning{Liu Z. Jiang C et al. }            
   \titlerunning{A New Approach of Data-driven Simulation and Its Application to Solar Active Region 12673 }  
   \maketitle

%
\section{Introduction}           
\label{sect:intro}

Solar eruptive activity significantly impacts space weather, with solar flares and coronal mass ejections (CMEs) being the primary events originating from solar active regions \citep[ARs,][]{Webb2012CoronalME}. These phenomena release accumulated free magnetic energy in the coronal magnetic field, resulting in radiation emission, accelerating and heating the surrounding plasma \citep{Forbes2000}. The energetic plasmas carrying the Sun's magnetic field may interfere with the Earth's magnetic field and spacecraft, potentially causing disruptive effects on human activities. Therefore, understanding the dynamic processes of these eruptive events is crucial for mitigating their impact.

It is widely acknowledged that the variations of coronal magnetic field breed these eruptions. Nonetheless, direct observations of the coronal magnetic field are extremely challenging \citep{Lin2004}. While it is feasible to infer the trajectories of coronal magnetic field lines from some coronal loops using extreme-ultraviolet (EUV) and X-ray imaging \citep{Brosius2006, Tomczyk2008, Liu2009}, obtaining detailed information remains elusive. The primary source of magnetic field data available to us predominantly originates from the photosphere. Consequently, numerical simulation emerges as the important tool for studying the coronal magnetic field. One prominent numerical approach, the Nonlinear Force-Free Field (NLFFF) model \citep{Schrijver2006, Metcalf2008}, derivs the 3D coronal magnetic field through extrapolation from photospheric vector magnetograms. This modeling technique, refined over several decades, has significantly enhanced our understanding of the 3D coronal magnetic field configuration \citep{Wiegelmann2008, Regnier2013, Guo2017, Wiegelmann2021}.

The magnetic field in the NLFFF model is in static equilibrium (in which the magnetic pressure-gradient force is balanced by the magnetic tension force), thus impeding detailed investigations of the triggering and driving processes of eruption events, during which the field evolves dynamically and the forces are significantly unbalanced. Consequently, an data-constrained and data-driven magnetohydrodynamics (MHD) model, utilizing vector magnetograms as boundary conditions, has been proposed to simulate the dynamic evolution of the coronal magnetic field. This time-independent, data-driven model adeptly simulates the gradual, quasi-static evolution of the coronal magnetic field \citep{Wu2006,Wu2009, Chitta2014}, the rapid progression of solar eruptions \citep{Jiang2013, Kliem2013,Guo2021}, and the transition from quasi-static to fast eruption phases \citep{Jiang2016NC, Jiang2016HD, Guo2019, Zhong2021}. According to the different levels of simplification made to the MHD equations, these models can be classified into three groups: the magnetofrictional model \citep{Cheung2015, Price2019, Yardley2021}, the zero-$\beta$ MHD model \citep{Inoue2018, Guo2021, Kaneko2021}, and the full MHD model \citep{Fan2011, Feng2012, Guo2024}, further detailed information about these models can be found in \citet{Jiang2022}.

On the other hand, according to the ways of implementing the boundary condition of the data-driven model, it can be subdivided into the magnetic field ($\mathbf{B}$) driven \citep{Wu2006, Feng2015, Jiang2016NC, Jiang2016HD, Feng2017}, the velocity ($\mathbf{V}$) driven \citep{Guo2019, Hayashi2019, Jiang2021MHD, Zhong2021, Wang2022, Wang2023} and the electric field ($\mathbf{E}$) driven \citep{Cheung2012, Hayashi2018, Price2019, Pomoell2019}. The $\mathbf{B}$ driven is to use the observed magnetic field directly at the bottom boundary of the model, which is very straightforward, but this introduces inputs the magnetic divergence errors from the boundary \citep{Toth2000}. The $\mathbf{V}$ driven is to use the photospheric flow that can be recovered from a time sequence of observed magnetograms, and this velocity-recovering technique is relatively mature, such as the local correlation tracking (LCT) technique \citep{November1988, Chae2001, Fisher2008}, the differential affine velocity stimator (DAVE) method \citep{Schuck2006, Schuck2008, Schuck2019}. The $\mathbf{V}$ driven model gives a good representation of the surface motions of magnetic flux, such as shearing and rotation of sunspots, but cannot be sufficiently self-consistent for the calculation of emergence flux, for the vertical component of the photospheric velocity is difficult to recover by the currently available methods such as the DAVE4VM. \citet{Jiang2021MHD} found that there is a build-up of magnetic fields at locations of small velocities around the active region, and that the simulated magnetograms differ more and more from the observations as time goes on. The $\mathbf{E}$ driven, which uses the time series of the electric field as the bottom boundary, has the advantage that the magnetic field satisfies its no-divergence condition, and the corresponding disadvantage that the inversion of the electric field is more difficult and complex \citep{Fisher2010, Fisher2012, Kazachenko2014, Fisher2015, Lumme2017, Fisher2020}.

Although direct observation of the electric field in the solar photosphere is feasible using the Stark effect \citep{Wien1916}, the limited sensitivity of measuring instruments poses significant challenges \citep{Moran1991}. A common approach involves indirectly deriving the electric field from the photospheric magnetic field. Typical practice is dividing the electric field into an inductive part and a non-inductive part. While the inductive component can be derived by solving Faraday's law, and the non-inductive component is specified using different assumptions \citep{Fisher2010, Cheung2012, Cheung2015, Lumme2017}: neglected the non-inductive part ($\nabla \psi = 0$, $\psi$ is the non-inductive potential), the emergence of a twist field ($\nabla^2 \psi = -U (\nabla \times \mathbf{B}) \cdot \mathbf{\hat{z}}$), and a uniform vortial motion ($\nabla^2 \psi = -\Omega B_z$). The free parameters $U$ and $\Omega$ in the latter two assumptions require additional velocity field for estimation, such as Dopplergrams or DAVE4VM velocity. \citet{Kazachenko2014} describes the PTD-Doppler-FLCT-Ideal (PDFI) method, which, as the name suggests, includes the PTD method (which stands for poloidal-toroidal decomposition first introduced by \citet{Chandrasekhar1961}), the Dopplergram velocity, the optical flow of the FLCT method and the ideal constraint ($\mathbf{E} \cdot \mathbf{B} = 0 $). The PTD method serves as the inductive component, while the remaining elements are considered non-inductive component. A detailed implementation of the PDFI method can be found in \citet{Fisher2020}. 

 Another relatively straightforward approach involves using FLCT or DAVE4VM code to process vector magnetograms to obtain the velocity field, followed deriving the electric field using ideal Ohm's law ($\mathbf{E} = -\mathbf{V} \times \mathbf{B}$). However, due to the large noise levels in weak magnetic field, velocity inversion is highly susceptible to outliers, potentially leading to the manifestation of unphysical phenomena \citep{Schuck2008}. Thus, correction procedures are necessary after inversion of the electric field.

In this paper, we present a new approach to correct the DAVE4VM-based electric field, and perform a new E-driven MHD simulation for one-day evolution of AR~12673 using the corrected electric field. The details of the electric field inversion are given in Section~\ref{sec:E}. The model setup and parameters of the model are described in Section~\ref{sect:Method}. The results of the simulation are shown in Section~\ref{sect:Result}, and we conclude in Section~\ref{sect:Conclution}.

\section{Electric field inversion }
\label{sec:E}

We use the Space-weather HMI Active Region Patches \citep[SHARP,][]{Bobra2014} data product which provides the AR’s vector magnetograms with time cadence of $\Delta t_{\rm C} = 12$~min and pixel (grid) size of $0.5$~arcsec. Before using the DAVE4VM code, we first rebin the data to a grid size of $\Delta x = \Delta y = 1$~arcsec to reduce the computing time, and any data gap is filled with a simple linear interpolation in time. The DAVE4VM needs inputs of spatial (in the horizontal direction) and temporal derivatives of the magnetic field and a window size. All the derivatives of magnetic field (as mentioned in this section) are calculated by central difference, for example, 
\begin{align}
(\dfrac{\partial B^{\rm O}_z}{\partial t})^{n_{\rm C}}_{i,j} =\dfrac{(B^{\rm O}_z)^{n_{\rm C}+1}_{i,j} - (B^{\rm O}_z)^{n_{\rm C}-1}_{i,j}  }{2 \Delta t_{\rm C}}, \notag \\ 
(\dfrac{\partial B^{\rm O}_z}{\partial x})^{n_{\rm C}}_{i,j} =\dfrac{(B^{\rm O}_z)^{n_{\rm C}}_{i+1,j} - (B^{\rm O}_z)^{n_{\rm C}}_{i-1,j}  }{2 \Delta x}, 
\end{align}
where $n_{\rm C}$ is the time index. The window size is chosen as $11$~arcsec.

From the DAVE4VM code and the vector magnetogram $\mathbf{B}^{\rm O} $, we obtain the vector velocity $\mathbf{v}^{\rm D} $ and thus the electric field by assuming that $\mathbf{E}^{\rm D}=-\mathbf{v}^{\rm D} \times \mathbf{B}^{\rm O}$. The electric field can be decomposed into two components, i.e., $\mathbf{E}^{\rm D}= \mathbf{E}_{h}^{\rm D} + E_z^{\rm D}$, where $h$ denotes the horizontal component and $z$ the vertical component,
\begin{align}
 \mathbf{E}_h^{\rm D} &= -(v_y^{\rm D}  B_z^{\rm O} - v_z^{\rm D}  B_y^{\rm O}, v_z^{\rm D}  B_x^{\rm O} - v_x^{\rm D}  B_z^{\rm O}) , \notag \\
 E_z^{\rm D} &= -(v_x^{\rm D} B_y^{\rm O} -v_y^{\rm D} B_x^{\rm O}).
\end{align}
However, using $\mathbf{E}^{\rm D}$ in the magnetic induction equation ($\partial \mathbf{B}^{\rm O}/\partial t = -\nabla \times \mathbf{E}^{\rm D}$) does not recover the evolution of $\mathbf{B}^{\rm O}$, even only the vertical component $B_z^{\rm O}$ ! Since the horizontal velocity ($v_x^{\rm D} , v_y^{\rm D} $) is often more accurate than the vertical velocity ($v_z^{\rm D} $), the vertical electric field $E_z^{\rm D}$ should be more accurate than the horizontal electric field $ \mathbf{E}_h^{\rm D}$, and therefore we have more freedom to the modify $\mathbf{E}_h^{\rm D}$. Our purpose is to modify the $\mathbf{E}^{\rm D}$ to a new electric field $\mathbf{E}$ such that using the magnetic induction equation the $B_z^{\rm O}$ can be fully recovered.

Before this, we smoothed the data $\mathbf{B}^{\rm O} $ and $\mathbf{E}^{\rm D}$ since in our data-driven model, the implementation of bottom boundary conditions is based on numerical difference. These data were spatially smoothed using a Gaussian smoothing with FWHM of $8$ arcsec, and temporally smoothed using a boxcar average of $120$ min, and the smoothed version of the two fields is also denoted by $\mathbf{B}^{\rm O} $ and $\mathbf{E}^{\rm D}$. 

To recover the evolution of $B_z^{\rm O}$, the horizontal electric field $ \mathbf{E}_h$ can be decomposed into two parts, 
\begin{equation}
\mathbf{E}_h = \mathbf{E}_h^{\rm I} + \mathbf{E}_h^{\rm N},
\end{equation}
where $\mathbf{E}_h^{\rm I}$ is the inductive part, $\mathbf{E}_h^{\rm I} = \nabla_h \times \phi \vec e_z = \left(\dfrac{\partial \phi }{\partial y}, -\dfrac{\partial \phi }{\partial x}\right) $, and $\mathbf{E}_h^{\rm N}$ is the non-inductive part, $\mathbf{E}_h^{\rm N}= \nabla_h \psi = \left(\dfrac{\partial \psi }{\partial x}, \dfrac{\partial \psi }{\partial y}\right) $. The two scalars $\phi(x,y)$ and $\psi(x,y)$ are functions of $x$ and $y$. Using the vertical component of the induction equation, we have 
\begin{equation}
\dfrac{\partial B_z^{\rm O} }{\partial t} = - \nabla_h \times \mathbf{E}_h = -\nabla_h \times \mathbf{E}_h^{\rm I} = \dfrac{\partial^2 \phi }{\partial x^2} + \dfrac{\partial^2 \phi }{\partial y^2},
\end{equation} 
and solving this 2D Poisson equation in a rectangle region $A = [0,L_x] \times [0, L_y]$ obtains $\phi$ and thus $\mathbf{E}_h^{\rm I}$. In the paper we used the procedure IMSL$\_$POISSON2D in IDL to solve the Poisson equation. Two choices may be used for the boundary conditions. If we use the Neumann boundary conditions such that at the $x=0$ and $x=L_x$ boundaries we have $E_y^{\rm I} =- \dfrac{\partial \phi }{\partial x} = 0$ and at the $y=0$ and $y=L_y$ boundaries $E_x^{\rm I} = \dfrac{\partial \phi }{\partial y} = 0$, which means that $\mathbf{E}_h^{\rm I}$ is perpendicular to the boundary lines. This requires that 
\begin{equation}
\int_A \dfrac{\partial B_z^{\rm O} }{\partial t} ds = - \int_C \mathbf{E}_h^{\rm I} \cdot dl =0,
\end{equation} 
(where $C$ is the boundary curve of $A$) which is, however, not generally satisfied by the magnetogram. 

If use the Dirichlet boundary conditions $\phi=0$ on the boundary $C$, at the $x=0$ and $x=L_x$ boundaries we have $E_x^{\rm I} = \dfrac{\partial \phi }{\partial y} = 0$ and at the $y=0$ and $y=L_y$ boundaries $E_y^{\rm I} = -\dfrac{\partial \phi }{\partial x} = 0$, which means that $\mathbf{E}_h^{\rm I}$ is parallel to the boundary lines.

Then we can specify the non-inductive part, by assuming that 
\begin{equation}
\nabla_h \cdot \mathbf{E}_h^{\rm N} = \nabla_h \cdot \mathbf{E}_h^{\rm D},
\end{equation}
which requires that $\dfrac{\partial^2 \psi }{\partial x^2} + \dfrac{\partial^2 \psi }{\partial y^2} = \nabla_h \cdot \mathbf{E}_h^{\rm D}$, and solving this 2D Poisson equation obtains $\psi$ and thus $ \mathbf{E}_h^{\rm N} $. Also, two types of boundary conditions can be used. If we use the Neumann boundary conditions, then $ \mathbf{E}_h^{\rm N} $ will be parallel to boundary lines and if use the Dirichlet boundary conditions, $ \mathbf{E}_h^{\rm N} $ will be perpendicular to boundary lines. A good choice is that the total field $ \mathbf{E}_h $ to be parallel to the boundary lines, and thus the Dirichlet boundary conditions should be used for $ \mathbf{E}_h^{\rm I} $ and the Neumann boundary conditions for $ \mathbf{E}_h^{\rm N} $.

By combining the two parts, we have the final electric field that can be used in the numerical code as
\begin{equation}
\mathbf{E}=(\mathbf{E}_h ^{\rm I} + \mathbf{E}_h^{\rm N}, E_z^{\rm D} ).
\end{equation}

\section{Model}
\label{sect:Method}

We carried out data-driven simulation using the DARE--MHD code \citep{Jiang2016NC}. This model has been used in numerous data-driven simulations for solar coronal evolution and eruptions \citep[e.g.][]{Jiang2021MHD,Wang2022, Jiang2022, Wang2023}. In this study, the simulation volume is a cubic box in Cartesian coordinates system, with size of $512 \text{~arcsec} \times 512 \text{~arcsec} \times 512 \text{~arcsec}$ (corresponding to around 368 Mm) and the zero origin point $(x,y,z) =(0,0,0)$ at the center of the bottom boundary. The computational volume is resolved by adaptive mesh refinement (AMR) grid in the simulation, in which the highest resolution is $1$~arcsec and the lowest resolution is $8$~arcsec. To save computing time, we multiplied the observed magnetic field by a factor of $0.025$ (therefore the maximal magnetic field $B_z$ in the model is around $55$~G initially), and increased the candence of the input maps by $68.6$ times. The other parameters utilized in the model are the same as those in \citet{Jiang2021MHD} except that the kinetic viscosity coefficient is set to a small value of $ \nu = 0.1 \Delta x^2/\Delta t$. In addition, all the variables in the numerical model are normalized using typical coronal values: $\rho_{\rm cor} =2.29 \times 10^{-15} $ g cm$^{-3}$ (density), $T_{\rm cor} = 1 \times 10^6 $ K (temperature), $H_{\rm cor} = 11.52 $ Mm (typical length in coronal scale), $B_{\rm cor} = 1.86 $ G (magnetic field), and $v_{\rm cor} = 110 $ km s$^{-1}$ (velocity). In the rest of the paper, all variables and quantities are mentioned as normalized values, if they are not specified.

\subsection{Initial conditions}

\begin{figure} 
   \centering
   \includegraphics[scale=1.3, angle=0]{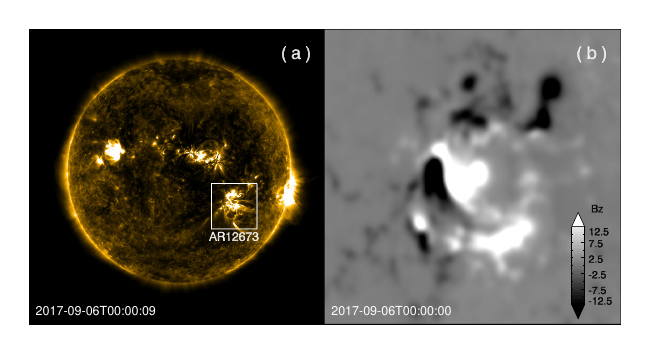}
   \caption{(a) The full solar disk image taken by SDO/AIA in EUV 171~{\AA}. The white box denotes the location of AR~12673. (b) The $B_z$ magnetogram of AR~12673 observed by the SDO/HMI at 2017 September 6 00:00~UT, with the resolution is $1$~arcsec and smoothed by Gaussian smoothing with FWHM of $8$~arcsec. }
   \label{ini}
   \end{figure}

AR~12673 is a very flare-productive region and has been studied by many authors~\citep[e.g.,][]{Yang2017, Chertok2018, Hou2018, Inoue2018, Liu2018, Jiang2018Magnet, Verma2018, Morosan2019, Romano2019, Zou2019, Zou2020, Guo2024}, delving into the intricate magnetic field structure of the AR and the mechanism of its eruptions. It appeared at the east solar limb on 2017 August 31, and disappeared on September 31, as observed by the Solar Dynamics Observatory (SDO). \Fig~\ref{ini}(a) shows the location of AR~12673 on the solar disk at September 6 00:00~UT. From September 4 to 10, there were four X-class flares and twenty-seven M-class flares observed, and two X-class flares occurred on September 6. The first X-class flare (X2.2) started at 08:57~UT, reached its peak at 09:10~UT, and ended at 09:17~UT. Subsequently, the second X-class flare (X9.3) began at 11:53~UT, peaked at 12:02~UT, and ended at 12:10~UT. Remarkably, the second flare marked the largest flare in solar cycle 24.

We used SDO/HMI vector magnetograms for AR~12673 in a one-day period from 2017 September 6 00:00~UT to 7 00:00~UT. The magnetograms have a resolution of $0.5$~arcsec and a cadence of $720$~s. The resolution used in our simulation is $1$~arcsec, which is rebinned from the original data in order to reduce the computational time. There is a data gap of magnetograms (from 06:00~UT to 08:36~UT) and we used a simple linear interpolation in time to fill the data gap. We chose the magnetogram of 00:00~UT on 2017 September 6 as the initial map and constructed the 3D magnetic field. Firstly, we smooth the vector magnetogram using Gaussian smoothing with FWHM of $8$~arcsec, as shown in \Fig~\ref{ini}(b), in order to filter out the small-scale structures and reduce the Lorentz force to make it easier to reach the equilibrium state. Then the initial 3D magnetic field is obtained by magnetic field extrapolation using the CESE--MHD--NLFFF code \citep{Jiang2013}. Finally, we input the extrapolated field to the DARE--MHD model, along with an isothermal plasma atmosphere of temperature $T = 1$ stratified by solar gravity with density $\rho = 1$ at the bottom boundary (same as \citet{Jiang2021MHD}), and relaxed them to an MHD equilibrium, which is used as the initial condition of subsequent data-driven simulation.

\subsection{Boundary conditions } 

  \begin{figure} 
   \centering
   \includegraphics[scale=0.8, angle=0]{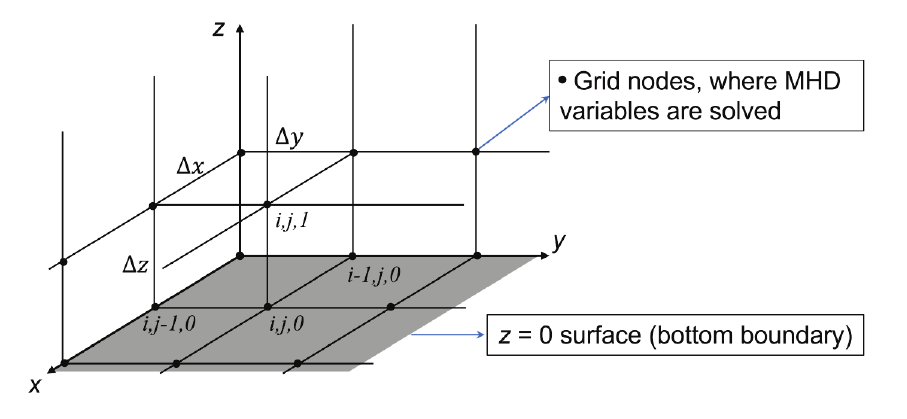}
   \caption{The grid structure near the bottom of the model.}
   \label{Fig:bottom}
   \end{figure}
Our model does not use ghost cell outside of the actual computational volume. The simulation volume extends from $z = 0$, and the $z = 0$ plane is exactly the bottom surface where the data-driven boundary conditions are applied.  In our code, all the MHD variables are assigned at the grid nodes (i.e., corner of the grid cells) rather than the cell center (as shown in \Fig~\ref{Fig:bottom}).

At the bottom surface, we update the magnetic field using the induction equation:
\begin{equation}
\label{eq:induc}
\dfrac{\partial \mathbf{B}}{\partial t} = -\nabla \times \mathbf{E} + \eta_{\rm stable} \nabla^2_{\bot} \mathbf{B}.
\end{equation} 
The  equation \ref{eq:induc} is discretized using a forward difference in time, a central difference in the horizontal direction on the surface, and a one-sided 2nd order difference in the z direction. In practice we found such a scheme could be unstable when $B_z$ is small, i.e., near the polarity inversion line (PIL). Therefore, a surface diffusion term $\eta_{\rm stable} \nabla^2_{\bot} \mathbf{B}$ is added to the induction equation, where $\nabla^2_{\bot} = \dfrac{\partial^2 }{\partial x^2} +\dfrac{\partial^2 }{\partial y^2}$ denotes the surface diffusion term, and the resistivity $\eta_{\rm stable} = 0.1 e^{-(B_z/4)^2}$ for numerical stability. Specifically, taking the $B_x$ component as an example, the scheme is
\begin{align}
\dfrac{(B_x)_{i,j,0}^{n+1}-(B_x)_{i,j,0}^n  }{\Delta t} = &- \dfrac{(E_z)_{i,j+1,0}^n - (E_z)_{i,j-1,0}^n }{2 \Delta y} + \dfrac{4(E_y)_{i,j,1}^n - 3(E_y)_{i,j,0}^n  -(E_y)_{i,j,2}^n }{2 \Delta z}  \notag \\
&+ \eta_{\rm stable} ( \dfrac{(B_x)_{i-1,j,0}^n-2(B_x)_{i,j,0}^n+(B_x)_{i+1,j,0}^n  }{\Delta x^2}  \notag \\
&+ \dfrac{(B_x)_{i,j-1,0}^n-2(B_x)_{i,j,0}^n+(B_x)_{i,j+1,0}^n  }{\Delta x^2} ),
\end{align}
where the subscripts $i,j,k$ denote the grid points in $x,y,z$ directions, respectively, and $k=0$ for the points at the bottom boundary. $\mathbf{E}_{i,j,0}$ is specified by the inversed electric field, $\mathbf{E}_{i,j,k}= - \mathbf{v}_{i,j,k}\times \mathbf{B}_{i,j,k}$ while for $k > 0$. 

With the magnetic field updated, we also need to update the plasma density, velocity, and pressure at the bottom boundary. Here the density and pressure are simply fixed as to be their initial values. Note that the electric field contains both the ideal and resistive part ($ \mathbf{E} = -\mathbf{v} \times \mathbf{B} +\eta \mathbf{j}$). If we assume that the current density is parallel to the magnetic field, $\mathbf{j} = \alpha \mathbf{B}$ (which is the force-free assumption), the velocity can be updated by: 
\begin{equation}
\label{eq:velocity}
\mathbf{v} = \dfrac{\mathbf{E} \times \mathbf{B}}{ B^2},
\end{equation}
since
\begin{equation}
-\mathbf{v} \times \mathbf{B}=-\dfrac{\mathbf{ E } \times \mathbf{B}}{B^2}\times \mathbf{B} = \dfrac{\mathbf{B} \times (\mathbf{ E } \times \mathbf{B})}{B^2} =\mathbf{ E } -\dfrac{(\mathbf{B}  \cdot \mathbf{ E }) \mathbf{B}}{B^2} = \mathbf{ E } -\dfrac{(\eta \mathbf{j}  \cdot \mathbf{B}) \mathbf{B}}{B^2} = \mathbf{ E } - \eta \mathbf{j} .
\end{equation}
In most cases, the force-free assumption is valid in the corona, thus it also applies the boundary surface (base of the corona). 

In addition, since the time step of the model $\Delta t$ is determined by the CFL condition, which is around $2$~s and is much smaller than inputting cadence of the electric field (which is $\Delta t_c /68.6 = 10.5$~s), interpolation of the electric field along time is required to provide the bottom-surface electric field needed by the model at each time step. We utilize a cubic spline interpolation scheme to maintain the continuity of the first-order time derivatives of the electric field (i.e., the changing rate of the electric field is continuous). In this way the changing rate of the magnetic field at the bottom surface is also continuous.

\section{Result}
\label{sect:Result}

\begin{figure}
   \centering
   \includegraphics[scale=0.8, angle=0]{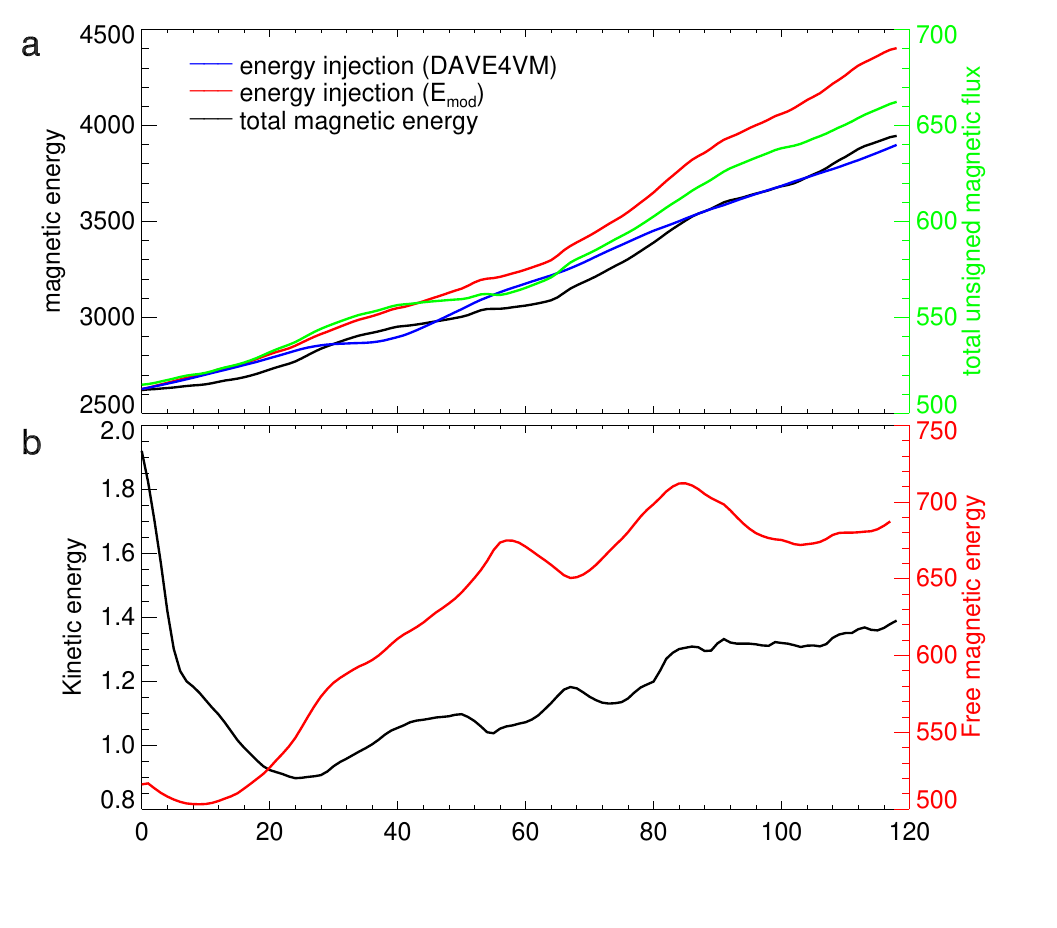}
   \caption{ (a) The evolution of the total unsigned magnetic flux (the green curve), the total magnetic energy (the black curve), the magnetic energy injection with the DAVE4VM (the blue curve) and simulated (the red curve), respectively. (b) The evolution of the free magnetic energy, and the kinetic energy.}
   \label{Fig:energy}
   \end{figure}

Firstly, we study the evolution of magnetic energy by comparing the volume-integrated value (i.e., magnetic energy stored in the corona) with the cumulative value from the bottom boundary (i.e., magnetic energy injected from the bottom boundary). This is an important way to check the consistence of the simulation. It is expected that, if without eruption, the magnetic energy stored in the corona should be close to (but less than, since there is energy dissipation in the corona) the energy injected from the bottom boundary. We calculated the Poynting flux through the bottom boundary as following,
\begin{align}
\label{eq:Einj}
P_{\rm inject} &=\int (\mathbf{E} \times \mathbf{B}) \cdot \mathbf{z} ~\rm d \textit{x} d \textit{y}.
\end{align}
and then integrated it over time as the magnetic energy injected into the corona. Besides, for comparison we also calculated the Poynting flux used directly the observed field $\vec B^{\rm O}$ and the DAVE4VM-derived electric field $\vec E^{\rm D}$, which are different from the values in our simulation. \Fig~\ref{Fig:energy}(a) presents the results. The blue and red curves represent the simulated and DAVE4VM-based magnetic energy injections, respectively, with the initial magnetic energy added up, while the black line depicts the variation of total magnetic energy in the model data-driven simulation. Overall, the three different values of magnetic energy increase in the simulated time interval, in agreement with the increase of the total unsigned magnetic flux (the green line). As expected, the magnetic energy injected from the bottom boundary in the simulation (i.e., the red line) exceeds slightly the volume-integrated magnetic energy (compare the red and black curves). Their deviation increases systematically with time, owing to numerical resistivity in the model. 

Initially, the magnetic energy injection in the simulation briefly aligns with the DAVE4VM-based magnetic energy injection. However, with time, the simulated magnetic energy injection begins to surpass that of the DAVE4VM-based injection. This difference continues to increase, reaching around $500$ (in normalized unit of energy) at the end of the simulation, constituting $40 \%$ of the DAVE4VM-based magnetic energy injection. Interestingly, the volume-integrated magnetic energy is rather close to the DAVE4VM-based energy injection (e.g., compare the blue and black curves). We note that the DAVE4VM-based energy injection (blue curve) exhibits a brief decrease during the data gap of the magnetograms (from $t = 30$ to $43$), which is not seen in the energy injection curve of the simulation.

\Fig~\ref{Fig:energy}(b) depicts the evolution of free magnetic energy (the red curve), and kinetic energy (the black curve). In the whole simulation the kinetic energy maintains very small values of below $10^{-3}$ of the magnetic energy, indicating that the system evolves in a quasi-static way. Therefore, the simulation did not reproduce the eruptions. Before $t=50$, the free magnetic energy and the total magnetic energy both exhibit a gradual increase. Following this period, the increase of the total magnetic energy experiences two brief decelerations, corresponding to the two decreases in free magnetic energy. Since no eruption is found in the simulation, the free energy decreases do not correspond to the two flares in observation.
   
\begin{figure}
	\centering
	\includegraphics[scale=0.40]{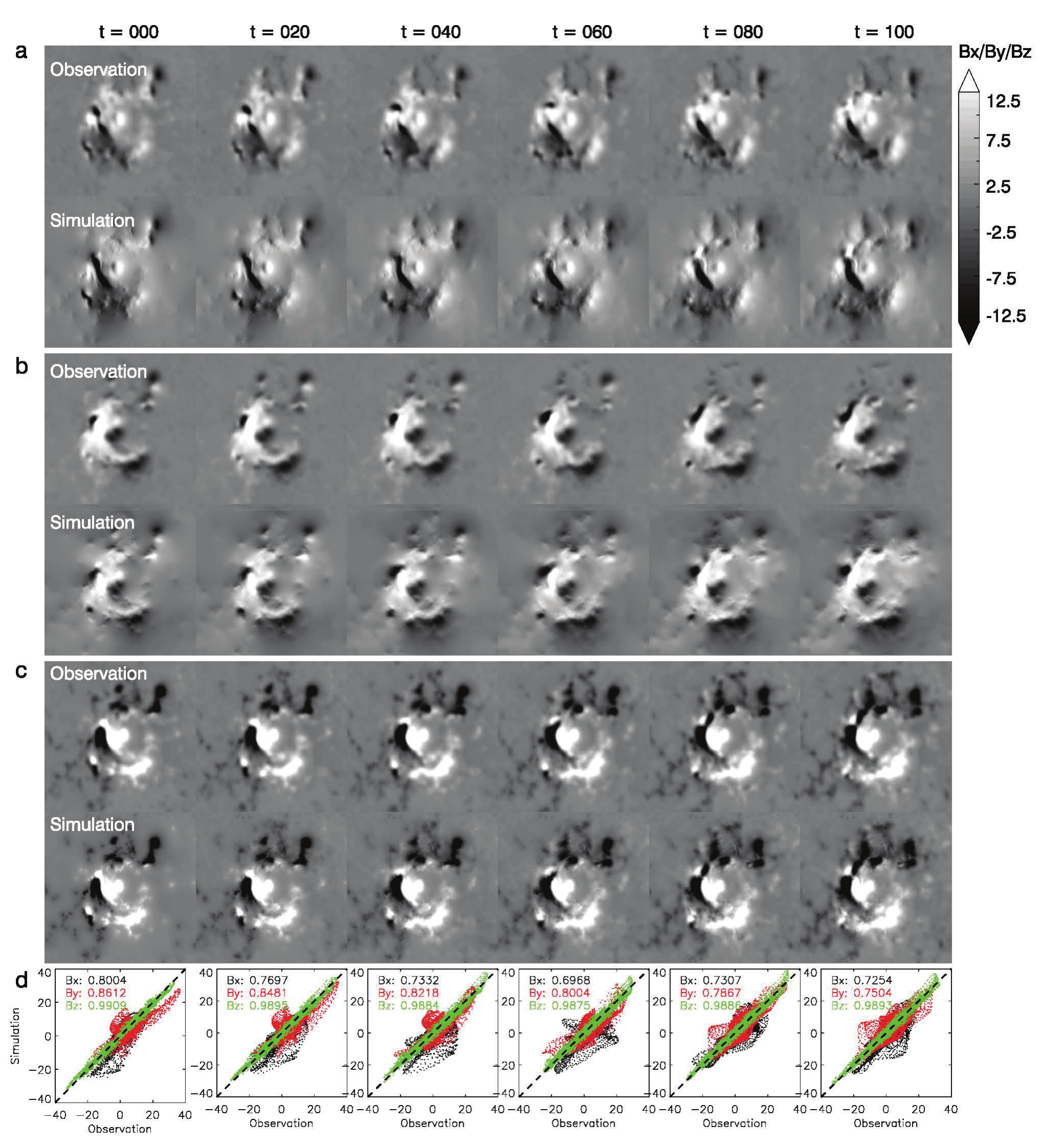}
	\caption{ (a-c) Comparison of observed and simulated $B_x$, $B_y$ and $B_z$, respectively. The observed magnetogram was multiplied by $0.025$ to match the simulation. (d) Scatter plots of the magnetic field at the bottom boundary of the model with the magnetograms, where black, red and green dots represent $B_x$, $B_y$ and $B_z$ values, respectively. The numbers indicate the Person correlation coefficients of the simulated values and magnetograms.}
	\label{Fig:map}
\end{figure}

We further look into how the simulated magnetic field at the bottom boundary differs from the observed data photospheric magnetic field. \Fig~\ref{Fig:map}(a-c) compares the magnetograms ($B_x$, $B_y$ and $B_z$, separately) in the simulation and the observation at different times. Overall, the distributions of the simulated field resemble the observed one throughout the entire evolution, except, some small-scale discrepancies. Unlike the velocity-driven simulation, the electric field-driven simulation does not exhibit any unreasonable significant magnetic field or pile-up at the edge of the main magnetic polarities (see \citet{Jiang2021MHD}).

\begin{figure}
	\centering
	\includegraphics[scale=0.8]{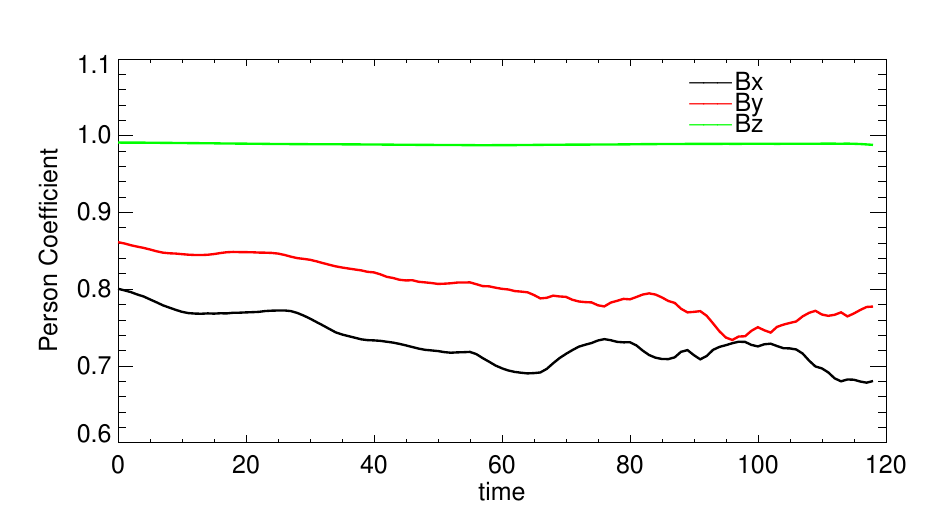}
	\caption{The evolution of the Pearson correlation coefficients of $B_x$ (the black line), $B_y$ (the red line) and $B_z$ (the green line).}
	\label{Fig:pearson}
	\end{figure}

Fig~\ref{Fig:map}(d) present the scatter plots of the magnetograms from both simulation and observation, where black, red and green dots represent $B_x$, $B_y$ and $B_z$ values, respectively. It is evident that the simulated $B_z$ values align very well with the magnetogram data, while $B_x$ and $B_y$ exhibit a slightly lower correspondence. The linear Pearson correlation coefficients are also shown in \Fig~\ref{Fig:map}(d), and their evolutions with time are shown in \Fig~\ref{Fig:pearson}. Note that our calculations of the correlation coefficients are based on the entire magnetogram and the full range of the magnetic field, which is different from some other authors who focus on particular regions and values~\citep{Price2019}. The results reveal Pearson correlation coefficients above $0.6$ for $B_x$, exceeding $0.7$ for $B_y$ and consistently approaching $1$ for $B_z$.

\begin{figure}
	\centering
	\includegraphics[scale=0.6]{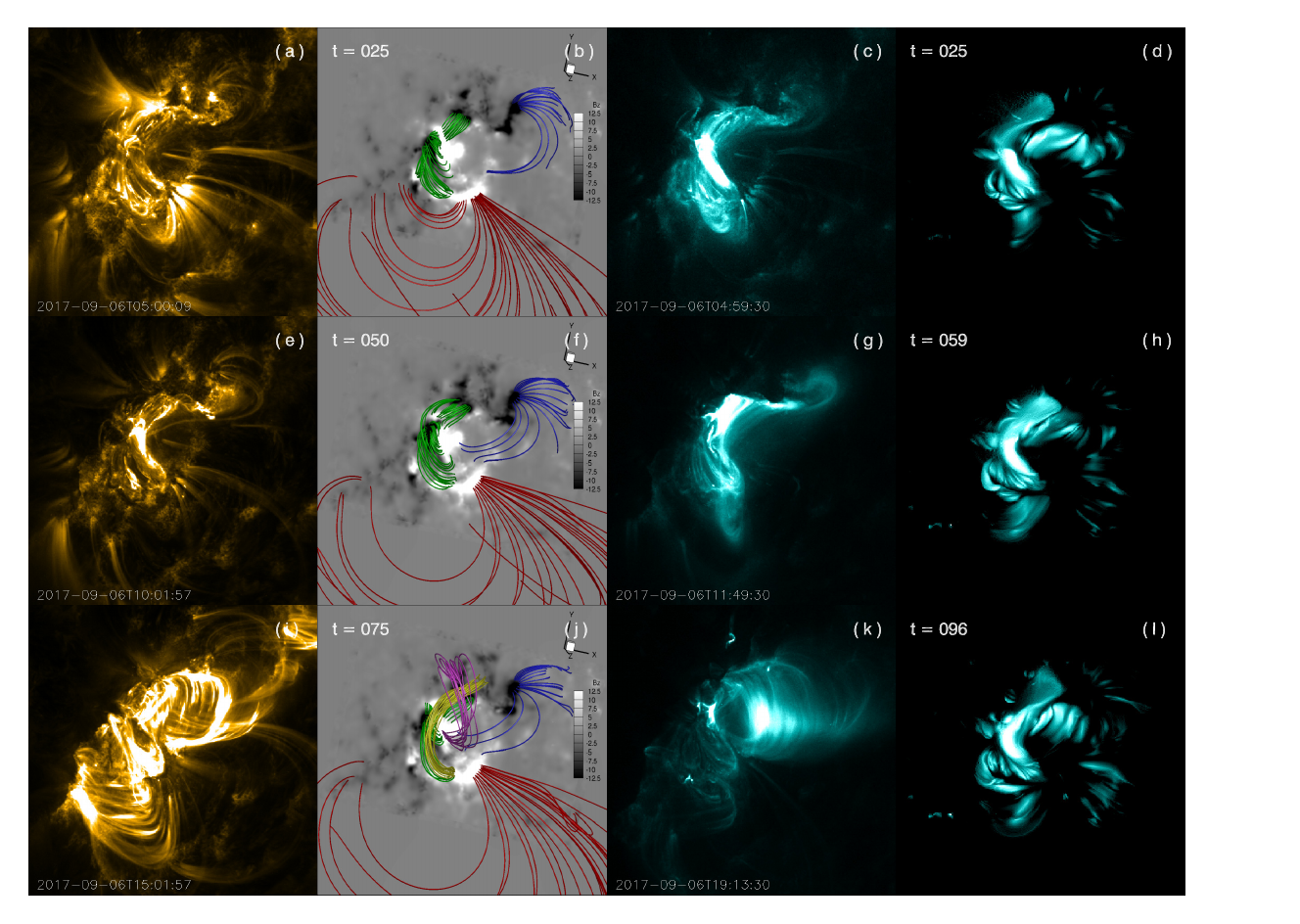}
	\caption{ The first (a, e, i) and third (c, g, k) columns are the SDO/AIA observed images in EUV wavelength of 171~{\AA} and 131~{\AA}, respectively. The second (b, f, j) column indicate the magnetic field lines at the corresponding moments, with different colors denote the different magnetic field lines, and the color of the background denotes the distribution of $B_z$. The fourth (d, h, l) column indicate the synthetic EUV image generated from the simulation results.}
	\label{Fig:AIA}
\end{figure}

Last, we study the coronal magnetic structure by comparing the EUV observed images with the simulated magnetic field in \Fig~\ref{Fig:AIA}. The first (a, e, i) and third (c, g, k) columns are the SDO/AIA observed images in EUV wavelength of 171~{\AA} and 131~{\AA}, respectively. Panels (b, f, j) are the magnetic field lines at the corresponding moments. Panels (d, h, l) are the synthetic images generated from the current density in the corona, following the method as proposed by~\citet{Cheung2012} and \citet{Jiang2016HD}. As can be seen, in the early stage of the simulation, a good morphological similarity is achieved between the simulated emission and the observed EUV images. However, after the X9.3 flare time (near $t=60$), the magnetic field structure diverges from observations due to the absence of eruption and its significant changes in the simulation. Nevertheless, our simulation shows the formation of a magnetic flux rope along the main PIL before the flare time, and such a C-shaped flux rope is also revealed in previous studies by~\citet{Jiang2018Magnet} and \citet{Guo2024} using different modeling techniques including NLFFF extrapolation and magneto-frictional model. After the flare time, this flux rope starts to rise gradually and reaches a higher position at the end of the simulation without showing impulsive eruption. As shown by the yellow lines in \Fig~\ref{Fig:AIA}(j), the flux rope still exists after the flare time at $t = 75$. There are several reasons why the flux rope did not erupt, for example, the magnetic free energy is not sufficient enough owing to the loss of energy by the numerical diffusion in the corona; the boundary conditions have not fully reproduced the transverse field of the magnetograms.

\section{Conclution and discussion}
\label{sect:Conclution}

In this paper, we first presented a new approach for photospheric electric field inversion by employing the photospheric velocity derived from vector magnetograms with the DAVE4VM code. With the photospheric velocity and magnetic field given, the electric field is simply derived based on Ohm's law, but the magnetic flux distribution cannot be recovered using the magnetic induction equation. This motivated us to modify the horizontal component of the derived electric field by decomposing it into an inductive part and a non-inductive part, where the inductive part is directly solved using the vertical component of the induction equation.

After obtaining the time series of the corrected electric field, we studied the magnetic dynamic evolution of AR~12673 from 2017 September 6 to 7 using a data-driven MHD model driven by the electric field. The result shows that the new approach enhances the magnetic energy injection through the bottom surface by $40 \%$ compared to that of the values by the original DAVE4VM-derived electric field. On the other hand, due to the energy dissipation in the corona, the volume-integrated magnetic energy in the corona is rather close to the DAVE4VM-based energy injection. The evolution of both total and free magnetic energy reflects the variation of in the bottom magnetic field. Therefore we compared the magnetic field at the bottom boundary in the simulation with the observed magnetograms. As expected, the evolution of the photospheric vertical magnetic field is well recovered with a correlation coefficient of essentially unity. The main structures of the horizontal components are recovered, and all the correlation coefficients are mainly above $0.7$. We also compared the simulated coronal magnetic field with EUV observations. At an early stage, the coronal magnetic field structure in our simulation exhibits a notable resemblance to EUV observations. However, at a later stage (after $t=60$) disparities emerge between the simulated magnetic field structure and the post-eruption observations. This discrepancy arises from the absence of eruptions phenomenon in the simulation, which precludes large-scale magnetic field reconnection and reconfiguration.

In future study, we will consider to use the more advanced electric field inversion method, e.g., the PDFI$\_$SS code~\citep{Fisher2020}, to provide an electric field that can hopefully recover all the three components of the photospheric magnetic field. Moreover, a high-accuracy simulation is required to reduce the magnetic energy loss due to the numerical resistivity, such that most of the energy as injected from the bottom boundary can be accumulated in the coronal field until an eruption is initiated.

\normalem
\begin{acknowledgements}
This work was jointly supported by National Natural Science Foundation of China (NSFC 42174200), Shenzhen Science and Technology Program (Grant No. RCJC20210609104422048), Shenzhen Key Laboratory Launching Project (No. ZDSYS20210702140800001), Guangdong Basic and Applied Basic Research Foundation (2023B1515040021), and the Fundamental Research Funds for the Central Universities (Grant No. HIT.OCEF.2023047). 

\end{acknowledgements}
  
\bibliographystyle{raa}
\bibliography{ED}

\end{document}